# Elucidating Dynamic Conductive State Changes in Amorphous Lithium Lanthanum Titanate for Resistive Switching Devices

Ryosuke Shimizu[1], Diyi Cheng[2], Guomin Zhu[1], Bing Han[1], Thomas S. Marchese[3], Randall Burger[1], Mingjie Xu[4], Xiaoqing Pan[4], Minghao Zhang[1,*], and Ying Shirley Meng[1,2,3,*]

[1]*Department of NanoEngineering, University of California San Diego, La Jolla, CA, USA.*
[2]*Materials Science and Engineering Program, University of California San Diego, La Jolla, CA, USA.*
[3]*Pritzker school of Molecular Engineering, University of Chicago, Chicago, IL, 60637, USA*
[4]*Irvine Materials Research Institute, University of California Irvine, Irvine, California 92697, USA*

* Co-correspondence: miz016@eng.ucsd.edu, shirleymeng@uchicago.edu

**Abstract:** Exploration of novel resistive switching materials attracts attention to replace conventional Si-based transistor and to achieve neuromorphic computing that can surpass the limit of the current Von-Neumann computing for the time of Internet of Things (IoT). Materials priorly used to serve as an electrode or electrolyte in batteries have demonstrated metal-insulator transitions upon an external electrical biasing due to resulting compositional change. This property is desirable for future resistive switching devices. Amorphous lithium lanthanum titanate (a-LLTO) was originally developed as a solid-state electrolyte with relatively high lithium ionic conductivity (~$10^{-3}$ S/cm) and low electronic conductivity (~$10^{-10}$ S/cm) among oxide-type solid electrolytes. However, it has been suggested that electric conductivity of a-LLTO changes depending on oxygen content. In this work, the investigation of switching behavior of a-LLTO sandwiched by nickel electrodes was conducted by employing a range of voltage sweep techniques, ultimately establishing a stable and optimal operating condition within the voltage window of -3.5 V to 3.5 V. This voltage range effectively balances the desirable trait of a substantial resistance change by three orders of magnitude with the imperative avoidance of LLTO decomposition. This switching behavior is also confirmed at nanodevice of Ni/LLTO/Ni through *in-situ* biasing inside focused-ion beam/scanning electron microscope (FIB-SEM). Experiment and computation with different LLTO composition shows that LLTO has two distinct conductivity states due to Ti reduction. The distribution of these two states is discussed using simplified binary model, implying the conductive filament growth during low resistance state. Consequently, our study deepens understanding of LLTO electronic properties and encourages the interdisciplinary application of battery materials for resistive switching devices.

## Introduction

Neuromorphic devices aim to mimic the speed of a human brain and would facilitate a revolutionary increase in computation speed. They resultingly attract attention to fulfil the needs of computationally intensive artificial intelligence (AI) and also complex pattern recognition and modelling for climate systems[1–4]. Resistive random-access memory (ReRAM) emulates synaptic behavior and is one of the promising memory devices owing to its simple metal-insulator-metal (MIM) structure, non-volatility, and low power consumption[3]. In ReRAM devices, the resistance of the material changes under an external electric field. In general, insulative materials used act as the high resistance state (HRS), or OFF state, at the beginning, which are then switched to low resistance state (LRS), or ON state, by applied electrical stimuli. So far, various oxide materials such as binary oxides (e.g. $HfO_x$[5–8], $AlO_x$[9–11], $NiO$[12–15], $TiO_x$[16–18], and $SiO_x$[19–22]) and ternary oxides (e.g. $SrTiO_3$[23–27]) have been explored as the switching material in a ReRAM device. The underlying mechanism of their switching behavior is proposed to be the formation of conductive filaments penetrating through the insulator, resulting in reduced resistance of the whole device[3,28,29]. However, a more comprehensive understanding of the resistive switching mechanisms is lacking for emerging smart materials.

In parallel to study of ReRAM devices, all solid-state thin-film battery is another intriguing field for both fundamental interface studies and application for micropower source[30–36]. Battery studies share common features of interest to ReRAM such as dynamic ion and defect migration ($Li^+$ and oxygen vacancy)[37–39], phase transformation under operation[40–43], and similar component architecture (electrode-electrolyte-electrode, **Figure S1**)[3,31]. These similarities enable synergetic investigations, toward deeper understanding of both fields by discoveries in one. Indeed, these days, knowledge acquired through battery studies are transferred to develop new resistive switching devices. For example, Fuller et al. explored $LiCoO_2$, a popular cathode material as a potential material for a neuromorphic computing device[42]. This is ascribed to an insulator-metal transition of $Li_xCoO_2$ ($0 \leq x \leq 1$) that changes its electronic conductivity by the factor of $10^6$ depending on lithiation state[44]. Young et al. reported that Li-ion battery anode material, $Li_xTi_5O_{12}$ ($4 \leq x \leq 7$), exhibits significant increase in electronic conductivity by a factor of ~$10^6$ upon only a few percent of lithiation[43]. Not only electrode materials can be used in a resistive switching device, but solid electrolyte materials are also ideal because of their electronically insulative pristine state as design.

Recently, our group has developed amorphous lithium lanthanum titanate ($Li_{3x}La_{2/3-x}TiO_3$, LLTO) thin film as a solid-state electrolyte in all solid-state thin-film batteries by using pulsed laser deposition[45]. LLTO possesses higher ionic conductivity (~$10^{-3}$ S/cm) than a conventional thin film solid state electrolyte, such as lithium phosphorous oxynitride (LiPON, ~$10^{-6}$ S/cm). In the study, we optimized thin film deposition pressure and temperature to maximize its ionic conductivity with sufficiently low electronic conductivity. Amorphous structure facilitates ionic conduction by eliminating grain boundaries in its resistance, which previously limited LLTO ionic conductivity below $10^{-5}$ S/cm.[46]

Beyond its application to energy storage devices, recent work on amorphous LLTO (a-LLTO) done by Shi et al. has demonstrated a memristive switching response under the stimuli of an external electric field, indicating the potential of integrating a-LLTO as a neuromorphic material for the future generation of memristors[47]. Deng et al. recently reported that LLTO shows switching behavior by the order of $10^2$.[48] They also claim rectifying effect in Pt/LLTO/Pt device. This is explained by the different types of metal-semiconductor contact at the top electrode/LLTO interface (Schottky contact) and LLTO/bottom electrode (quasi-ohmic contact), allowing current flow in one specific direction. However, composition of thin film LLTO is quite far from the original design (about 10% of Li from original LLTO stoichiometry) and there is little of the direct evidence about such interface contact differences. Furthermore, they tested the Pt/LLTO/Pt device up to 6 V, which is too high voltage for reversible stable cycling behavior. Therefore, the resistive switching mechanism of a-LLTO including long-term stability is yet to be fully elucidated. Although the resistive switching mechanism is potentially supported by the formation of conductive pathway inside a-LLTO, lack of characterization method to identify the growth of such a conductive pathway makes it inconclusive. Detailed characterization has thus far not been possible due to the amorphous nature of a-LLTO (i.e., no crystalline ordered structure) and nanoscale filament growth.

In this work, we demonstrate successful observation of LLTO resistive switching behavior and optimization of the voltage range to achieve both resistive switching and cycling stability. The resistance change is also demonstrated in LLTO nanodevice, which is prepared and tested in focused ion beam (FIB)-scanning electron microscope (SEM). Furthermore, electronic conductivity changes that provoke the LLTO switching are investigated through combined experimental approach and first-principles analysis. Drawing insights from both experimental findings and computational data, we engage in a comprehensive exploration of switching mechanisms within the context of a simplified LLTO model. We propose the metastable conductive LLTO state formed by local compositional change under biasing bridges one side to the other side of electrode, resulting in resistive switching in the device.

**Results and discussion**
*Switching behavior of Ni/a-LLTO/Ni device*
LLTO thin film was prepared by pulsed laser deposition (PLD) under the conditions that our group previously optimized[21]. Details are illustrated in the experimental section. The deposited film shows amorphous characteristics and columnar structure as shown in **Figure S2**. The switching behavior of a-LLTO with Ni electrodes was evaluated when cycled from 0 V→ 3.5 V→ -3.5 V→ 0 V. The device transitioned from HRS to LRS from 0 V to 3.5 V and becomes LRS to HRS from 3.5 V to 0 V (**Figure 1 (a)**). The corresponding current difference at 1.0 V, for instance, spanned several orders of magnitude (from $8\times10^{-8}$ at HRS to $5\times10^{-6}$ A at LRS), large enough for operation as a resistive switching device. **Figure 1 (b)** shows resistance vs. voltage plot of the device, also suggesting resistance decreases during the sweep 0 V→ 3.5 V, i.e., ON, and increases during the sweep 0 V→ 3.5 V, i.e., OFF. To identify the optimal voltage range, different voltage ranges were then explored and compared in **Figure 1 (c)**, where the ratio of end-point current to current at 1.0 V as a reference point. Here, end-point current is defined as the current value at the maximum voltage in each voltage sweep. For example, when the device was cycled at the range of -3.5 V ≤ V ≤ 3.5 V, end-point current is the value at 3.5 V. When the voltage range is -3.0 V ≤ V ≤ 3.0 V or smaller, the magnitude of current ratio is on the order of $10^2$ while when the voltage range is -3.5 V ≤ V ≤ 3.5 V or larger it reaches $10^3$. In **Figure S3** and **S4**, I-V curves at different voltage range are demonstrated as linear-linear and log-linear plot, respectively. At lower voltage range (e.g., -2.5 V ≤ V ≤ 2.5 V), its I-V curve is like "butterfly-shape" and different from the one at higher voltage range shown in **Figure 1 (a)**. Similar I-V behavior was observed in the study by Deng et al.[48] and Kim et al.[49], implying the resistive switching is not activated at this lower voltage range. Thus, these results show that the device needs a certain amount of voltage to turn on the resistive switching, i.e., around 3.5 V in the above case.

The durability of the devices cycled between different voltage ranges was studied as well. In **Figure S5**, the corresponding end-point current was plotted as a function of cycling numbers. **Figure S5** shows that when the device is cycled at high voltage range such as 4.5 V and 5.0 V, the current change over cycling is significant. This is because the voltage range goes above the voltage stability window of LLTO. Zhu et al. calculated the electrochemical window by density functional theory (DFT) that is within 1.8 - 3.7 V.[50] Therefore, the device is not stably cycled across a higher voltage range, even though it shows resistive switching behavior. In **Figure S6**, TEM images of a-LLTO before and after biasing to 5.0 V are displayed. While pristine a-LLTO shows amorphous feature in the magnified image and its FFT pattern (**Figure S6 (c)**), in the a-LLTO after biasing, a periodic feature is partially observed, as shown in **Figure S6 (e) and (f)**. This feature in FFT pattern has 2.98 Å spacing and equivalent to the crystalline of $La_2Ti_2O_7$,[51] which supports calculations that high voltage operation can cause the decomposition of a-LLTO.

In summary of the discussion on operation conditions of the voltage range, within the voltage stability window of a-LLTO, a higher voltage than 3.0 V should be selected to make the difference of the current at HRS and LRS as distinguishable as possible. If the voltage range is too small, the device does not show appreciable switching behavior, as we discussed above. If the operating voltage is too high, such as 4.5 V or higher, a-LLTO is decomposed. As such, in order to retain distinguishable resistive switching with relatively good cycling stability of the device, a voltage range of $-3.5\ V \leq V \leq 3.5\ V$ should be used for the a-LLTO device.

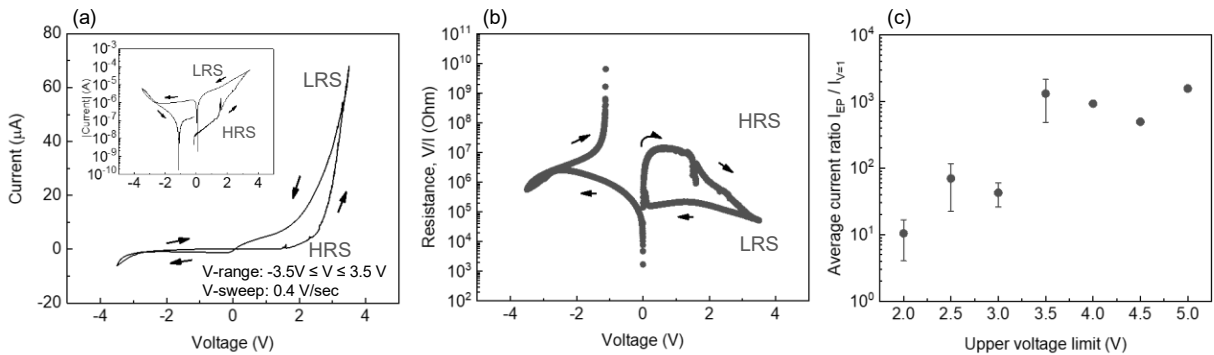

**Figure 1. Electrical response of Ni/LLTO/Ni thin film during voltage sweep.** (a) Linear I-V plot and log |I|-V plot (inset) of LLTO, where voltage was swept between -3.5 and 3.5 V. The voltage was swept at the rate of 0.4 V/sec. (b) Resistance (V/I) vs. V plot, when voltage was swept between -3.5 and 3.5 V. (c) Average ratio of current at the highest voltage in the different voltage range, $I_{EP}$ (EP: end-point), to current at 1.0 V, $I_{V=1}$, in the first cycle. For example, when voltage was swept between -3.5 and 3.5 V, upper voltage limit is 3.5 V and $I^{EP}$ is the current value at 3.5 V.

In addition to the bulk scale, we also demonstrate a nano scale device with a similar resistance change. A specially designed chip was used to measure nano scale device's electrical response inside a FIB-SEM. Nanodevice preparation is depicted in **Figure S7 (a)**, where a lamella consisting of Ni/LLTO/Ni layers was lifted out from thin film device and mounted onto the biasing chip. **Figure S7 (b)** demonstrates the final configuration after nanodevice preparation. **Figure S7 (c)** shows voltage response of the device under each constant current. The voltage response of the nanodevice is similar to the bulk device results obtained by voltage sweeping (**Figure S7 (c)**). Furthermore, resistance change is visualized by plotting the differential resistance as a function of the voltage in **Figure S7 (d)**, suggesting that the resistance of the device experiences a dramatic decrease between 0.1 V and 1 V. This result clearly implies that, as well as the bulk-scale, the nano device turns into a "ON" state by applied high current/voltage and proves a potential for the LLTO device to operate similarly to a conventional transistor.

*Effect of oxygen vacancy on electronic conductivity of LLTO*
When a-LLTO is under an electrical bias, $O^{2-}$ ions migrated towards the positive electrode due to Coulombic attraction. The microscopic composition of LLTO under biasing is speculated to show gradational change across the thickness direction, in which LLTO becomes O-rich on positive side and O-poor on negative side (**Figure 2 (a)**). To explore the effect of the different oxygen composition on the electronic conductivity, a-LLTO thin films were deposited at various $O_2$ partial pressures during PLD (**Figure 2 (b)**). There appears to be a critical $O_2$ pressure around 0.08 mbar, where the electronic conductivity of a-LLTO film changed drastically by more than five orders of magnitude. LLTO films deposited above this critical $O_2$ pressure behave as insulators. The electronically conductive film under this point is attributed to the introduction of oxygen vacancy into a-LLTO (O-poor LLTO). Our past study by Lee et al. discussed that when oxygen vacancies form in LLTO, $Ti^{4+}$ cations are reduced to $Ti^{3+}$ for the charge compensation, and, consequently, an electron conduction pathway is formed[45].

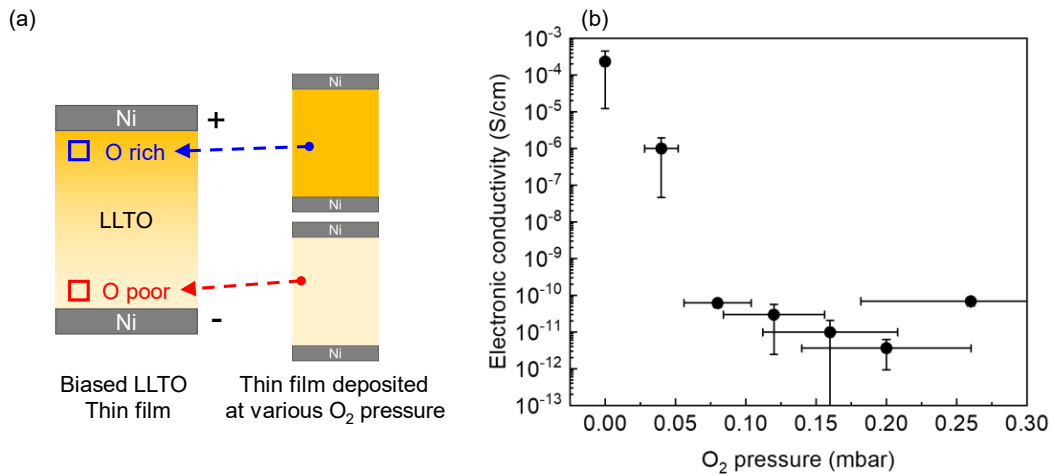

**Figure 2. Electrical conductivity dependency of a-LLTO on oxygen content.** (a) Schematic illustration of (left) biased a-LLTO and (right) thin film deposited at various $O_2$ partial pressure to demonstrate local composition in biased LLTO film. (b) Electronic conductivity vs. $O_2$ deposition pressure during PLD. The horizontal error bar is attributed to the accuracy of the pressure gauge, which is ±30%. The vertical error bar is from statistics of multiple measuring points on one sample.

When oxygen ions migrate to the positive side upon biasing in a-LLTO, lithium ion migrate to the negative side. DFT calculations were employed to simulate the changing electronic structure of LLTO upon different compositions. Here we employ crystalline LLTO to simplify calculations. Starting with cubic LLTO structure with 4 Li atoms, 4 La atoms, 8 Ti atoms, and 24 O atoms (**Figure 3 (a)**), LLTO composition was changed to simulate various Li and O contents present within. Three different O compositions that contain 24 O atoms, 23 O atoms and 22 O atoms were simulated, corresponding to 0%, ~4.2 %, and ~8.3 % of O vacancy, respectively. Similarly, the number of Li atoms was varied from 0 to 4 in the LLTO structure for a total of fifteen cases. **Figure 3 (c)** summarizes the band gap values of LLTO across fifteen compositions obtained from each electronic structure (e.g., **Figure 3 (b)**), where LLTO shows two major categories – the conductive phase (green) and insulative phase (red). In the structures where Li content is stoichiometric or deficient, LLTO band gaps decrease with increased oxygen vacancy density, indicating a rise of electronic conductivity. This has a good agreement with the experimental results of the electronic conductivities of LLTO film deposited at different $O_2$ pressures in **Figure 2** and suggests that oxygen vacancy could serve as the key parameter for induced conductivity change and resistive switching behavior. As exemplified $TiO_2$, $TiO_{2-x}$, and $Ti_2O_3$, presence of oxygen vacancy triggers Ti reduction, reducing the band gap[21, 27–30]. In $TiO_2$, all Ti is theoretically $Ti^{4+}$ ($3d^0$) and a band gap exists between filled O 2p orbital and empty Ti 3d orbitals. Once an oxygen vacancy is introduced to the system ($TiO_{2-x}$), oxygen vacancy state is created between the valence and conduction band. When Ti is fully reduced to $Ti^{3+}$ ($3d^1$) in $Ti_2O_3$, Ti 3d orbitals become partially occupied and create a 0.1 eV gap due to the strong repulsion force among those $3d^1$ electrons. In the case of LLTO, many researches aiming for LLTO as an electrolyte materials in contact with Li metal anode prove that Li insertion into LLTO can be the other cause that induces Ti reduction[55–57]. Such study shows once LLTO is in contact with Li metal, Ti is spontaneously reduced from $Ti^{4+}$ to $Ti^{3+}$, which changes its electrical properties. In the structure that contains a stoichiometric 23 O atoms, the band gaps experience an increase and subsequent decrease as Li content is reduced, implying that Li composition also influences the electronic structure of LLTO.

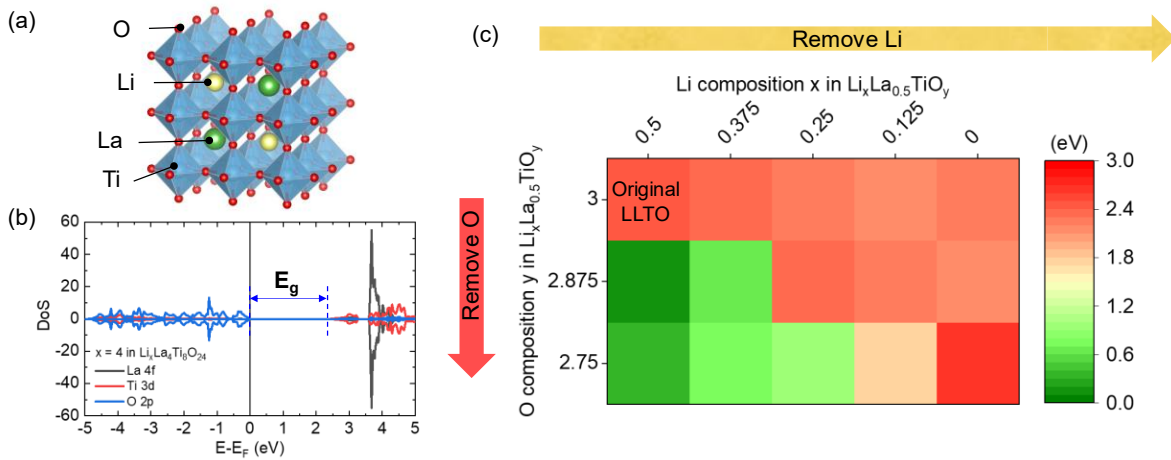

**Figure 3. Computational analysis on band gap of LLTO with different Li and O composition.** (a) Structure of $Li_4La_4Ti_8O_{24}$ that was used for DFT calculations. (b) Electronic structure of pristine LLTO ($Li_4La_4Ti_8O_{24}$). (c) Summary of LLTO band gaps with different Li and O compositions, in which the color code from green to red represents transition from conductive electronic properties to insulative.

*Resistive switching mechanism of a-LLTO*

Through experimental and computational approaches, LLTO is demonstrated to have two states (conductive and insulative) with different conductivities that trigger several orders of resistance change in the device during biasing. Here we discuss how those two states are anticipated to be distributed to affect conductivity changes, using a macroscopic and simplified binary system. Schematics in **Figure 4** depict two different scenarios of electronic conductivity change: a series and parallel connections of resistors representing variations of the two states in LLTO. If the conductive and insulative states are separately distributed across the thickness such as **Figure 4 (a)**, which is equivalent to a series connection, dramatical conductivity change cannot occur unless the conductive layer become over 95% of the whole film. No obvious morphological change was observed in nanodevice, not supporting the series connection scenario. In contrast, **Figure 4 (b)** simulates conductive state that forms a filament, bridging from one electrode to the other. In such a configuration, several orders of magnitude difference in electronic conductivity can be achieved through minor changes in filament size. According to **Figure 4 (b)**, filament diameter (in total) is expected to be ~10 μm in diameter to have $10^2$ order difference in the conductivity, assuming total area of the top electrode is 1 mm in diameter. This simple calculation offers guidance towards a future experimental approach, such as *operando* TEM or atomic force microscopy (AFM) with high special resolution, to identify filament growth for the next-stage investigation.

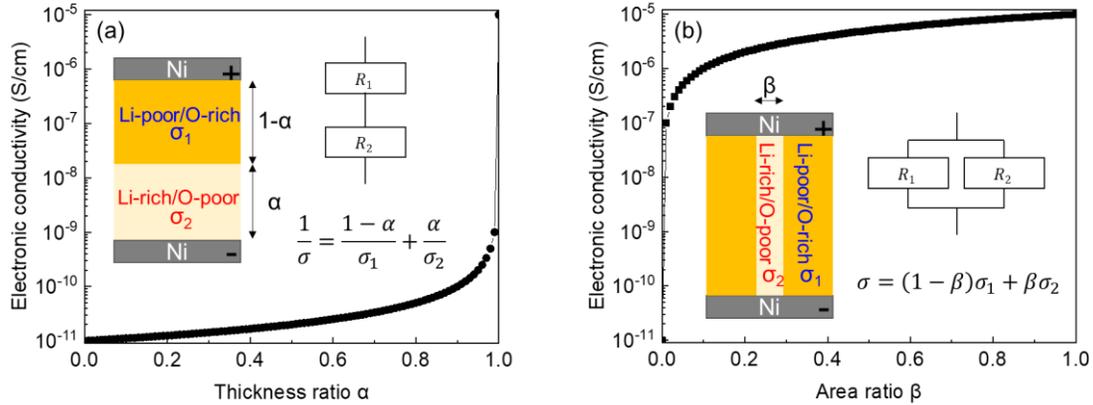

**Figure 4. Thought experiment to explain resistive switching mechanism with simple binary system.** (a) Scenario 1: two LLTO states exist as a serial circuit, where α represents thickness ratio of conductive (Li-rich/O-poor) state to the whole thickness. (b) Scenario 2: two LLTO states exist as a parallel circuit, where β represents cross-sectional area ratio of conductive (Li-rich/O-poor) state to the whole cross-section area. The inset for each figure illustrates the assumed system, equivalent circuit, and formula of combined conductivity. Here, electronic conductivity of Li-poor/O-rich state $\sigma_1$ and that of Li-rich/O-poor state $\sigma_2$ are $1 \times 10^{-11}$ S/cm and $1 \times 10^{-5}$ S/cm, respectively, which is based on experimental data in **Figure 2**.

Through the simplified calculation, it is found that the filament growth is critical to cause resistance jump by several orders of magnitudes under electrical biasing. Herein, we further investigate how a filament could grow inside LLTO. When LLTO is biased, charged mobile species, i.e., lithium ions and oxygen vacancies, migrate to the negative electrode. The migrated oxygen vacancies will be accumulated (**Figure 5 (b)**) and change local composition to form the conductive state of LLTO, in which electrons can freely move inside. At the conductive tip, electrons from conductive LLTO and oxygen vacancies from insulative LLTO meet to form another oxygen vacant LLTO under the reaction (**Figure 5 (d) right**), where LLTO loses oxygen and Ti is partially reduced from 4+ to 3+ for charge compensation. Thus, a new conductive layer of LLTO forms. By repeating this reaction, conductive filament extends toward the other electrode (**Figure 5 (c)**). The voltage acting as an external driving force must be sufficient to migrate charged species to nucleate large number of filaments to finally reach the other side, otherwise switching will not be achieved. In addition, oxygen vacancy formation energy calculation[58] in **Figure S8** suggests that O-deficient LLTO should not be energetically favorable. As demonstrated in **Figure S9**, DC bias testing implies that the ON state is spontaneously relaxed to an off state after releasing the switching voltage. These together explain that the filament does not remain after releasing voltage because of ion relaxation.

As battery materials, electrolytes have been developed to maximize properties as an ionic conductor and electronic insulator while cathode and anode materials are desirably mixed ionic-electronic conductor (MIEC). Thus, these materials have a strong potential for application as resistive switching active materials, which is triggered by ion migration. Based on the discussion so far, some guidelines for application of battery materials to future resistive switching devices can be proposed. First, materials should undergo electronic conductivity change by effect of local ion composition such as lithium or oxygen vacancy. Ionic conductivity plays a pivotal role here in determination of the switching speed. Another point evaluated in this study is stability. The material should have a metastable state or another phase to keep the conductive filament long even after removing external stimuli to make a non-volatile device.

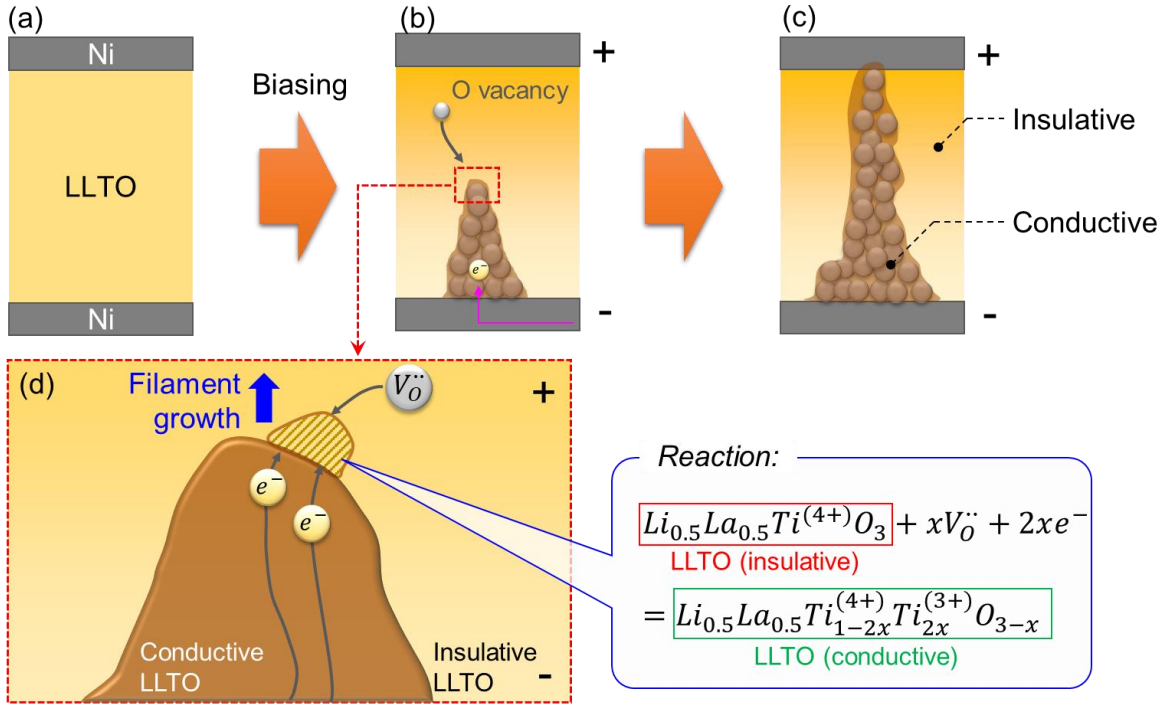

**Figure 5. Schematic illustration of filament growth inside LLTO.** (a) Pristine LLTO where all the mobile ion components (Li and O vacancy) are equally distributed. (b) Initial state of biased LLTO, where conductive filament starts to grow from negative electrode. (c) Biased LLTO, where conductive filament connects from negative to positive side of electrodes and lowers the total conductivity. (d) Magnified image of conductive filament tip in (b). On the tip of the filament, electrons can reach through the filament and oxygen vacancies are collected through the insulative LLTO state. The reaction occurs as expected (right) to form conductive LLTO, leading to growth of the filament.

**Conclusion**

In this work, we have confirmed that Ni/LLTO/Ni demonstrates the resistive switching behavior, where significant current changes by three orders of magnitude happen during voltage sweeping. Reversible witching requires optimization of voltage ranges. Too high voltage (4.5 V or above) causes LLTO decomposition, resulting in the degradation of the device. In contrast, too low voltage (3.0 V or below) is insufficient to activate the switching behavior. The electronic conductivity increase is attributed to the gradational composition difference inside LLTO. Oxygen vacancy mobility plays a pivotal role in device electronic conductivity, according to the experimental result that O-deficient LLTO becomes highly conductive due to the reduction of $Ti^{4+}$ to $Ti^{3+}$. Computational study exploring electronic structure of LLTO with varied Li and O compositions supports the experimental results. Combined the experimental and computational results led to a thought experiment to calculate the combined electronic conductivity of conductive and insulative LLTO states across two scenarios, series and parallel connection of the states. The parallel connection of states provides further ability to consider the process of "filament growth" inside the resistive switching device, supporting the several orders of magnitude change in resistance seen under electrical biasing.

**Experimental**

*LLTO target synthesis*

$Li_{0.5}La_{0.5}TiO_3$ was synthesized through solid-state reaction, based on past study[45]. A stoichiometric amount of $Li_2CO_3$ (Sigma Aldrich), $La_2O_3$ (Sigma Aldrich), and $TiO_2$ (anatase, Aldrich) powders were thoroughly mixed by a high energy ball mill (Emax, Retsch) in ethanol. Before mixing powders, $La_2O_3$ was dehydrated at 900°C for 4 hours with a ramping and cooling rate at 5°C/min in a box furnace. After ball-milling and removing ethanol, the mixed powder was calcinated in an alumina crucible in a box furnace. This calcination was conducted at 1200°C for 6 hours with a ramping and cooling rate at 5°C/min. The calcinated powder was grounded in a mortar and pestle and pelletized into a 1-1/8-inch die at 10 tons for 5 minutes. The pellet was sintered at 1300°C for 5 hours with the ramping and cooling rates at 5 °C/min. After sintering, the pellet surface was polished with sandpaper.

The molar ratio of Ti to La in the synthesized target was studied by electron energy-dispersive X-ray spectroscopy (EDS) equipped with FEI Quanta FEG 250, showing 2.12±0.37 (Designed value: 2). The target was fixed on the holder for pulsed laser deposition (PLD) processing.

*LLTO thin-film deposition and device fabrication*

With the synthesized target, LLTO thin film was deposited by PLD. The deposition was conducted by 248 nm KrF Lambda Physik LPX-Pro 210 excimer laser in an Excel Instruments PLD-STD-12 chamber. The laser was set at an energy fluence ~2 J/cm$^2$ and 4 Hz frequency for 5000 shots during the deposition. Ni-coated SiO$_2$/Si wafer (MTI corporation) was used as a substrate, and the LLTO film was grown at 400°C substrate temperature under 200 mTorr of O$_2$ pressure. On top of the LLTO film, 9 spots of Ni current collector (~ϕ1 mm) were deposited by DC sputtering (Denton Discovery 18 sputter system, National Nanotechnology Coordinated Infrastructure).

*Grazing Incidence Angle X-ray Diffraction (GIXRD)*
XRD pattern of the LLTO thin film was taken by Rigaku Smartlab X-ray diffractometer with Cu Kα source (λ = 1.5406 Å) with a working voltage and current of 40 kV and 44 mA, respectively, and a scan step size of 0.04°. The scan speed was 0.12° min$^{-1}$, and the scan range was from 0° to 80°.

*Electrochemical testing*
The electrochemical property of the deposited Ni/LLTO/Ni device was tested by a potentiostat with ultra-low current option, Biologic SP-200. Switching property was measured through the swept voltage at the rate of 0.04 V/sec with different voltage windows. Electronic conductivities of the films were also measured, where current was measured by applying constant voltage at 10 mV. Then, the electronic conductivity, $\sigma_e$, was calculated through the following equation:

$$\sigma_e = \frac{\ell I}{AV}$$

Here, $\ell, A, V,$ and $I$ are thickness of the film, area of top current collector, the applied voltage (10 mV), and measured current after relaxation, respectively.

*LLTO nanodevice fabrication and electrical biasing testing*
A FEI Scios DualBeam FIB/SEM equipped was used to mill and shape a lamella from bulk Ni/LLTO/Ni device. The operating voltage of the electron beam was 5 kV and emission current of the beam was 100 pA. A gallium ion beam source was used to mill, clean, and thin the sample with an operating voltage of 30 kV. Various emission currents were selected depending on purposes: 10 pA for ion beam imaging, 0.5 nA for cross-section surface cleaning and lamella thinning, and 5 nA for pattern milling. The fabricated lamella was mounted on the biasing chip (Hummingbird Scientific). The electrical response of the nanodevice was measured with Biologic SP-200 potentiostat through the voltage under different constant current sets which were applied on the chip via the EBIC port equipped outside of the FEI Scios DualBeam SEM. Here current, not voltage, was controlled to avoid the breakdown of nanodevice due to instant large current flow.

*Transmission electron microscopy (TEM) for pristine and biased LLTO*
The TEM specimen consisting of Ni/LLTO/Ni is prepared by the same procedure as mentioned above and mounted on an E-chip (E-FEF01-A4, Protochips). Thickness of the lamella was ≈500 nm while the center was thinned to ≈100-150 nm thick as a TEM window. TEM measurement was conducted with a Fusion 500 biasing holder (Protochips) on a JEOL 2800 TEM operating at 200 keV. The biased images were captured after voltage sweep at the rate of 2.0 mV/min from 0 to 5 V on the holder, using SP-200 Biologic equipped with a ultra-low current cable. Collected images were analyzed via DigitalMicrograph from Gatan.

*First-principles calculation*

In order to explain the effect of ions (Li and O) on LLTO, the electronic structure of LLTO was investigated in the spin polarized GGA+U approximation to Density Functional Theory (DFT)[59]. Projector augmented-wave method (PAW) pseudopotentials were employed as implemented in the Vienna Ab initio Simulation Package (VASP). The Perdew-Bruke-Ernzerhof (PBE) exchange correlation and a plane-wave representation for the wavefunction[60] were used, where a cut-off energy was set at 600 eV. The Brillouin zone was sampled with a k-points mesh of 3×3×3 for density of states (DOS) calculations. Effective U values used through all the calculations were 5 for Ti. Cubic spinel structure (space group: $Pm\bar{3}m$), $Li_4La_4Ti_8O_{24}$, drawn by VESTA[61] was used as an initial structure and the electronic structures of LLTO with different Li and O compositions were studied to simulate each point inside biased LLTO.

**CRediT authorship contribution statement**
**R.S.**: Conceptualization, Methodology, Software, Validation, Formal analysis, Investigation, Data curation, Visualization, Writing – original draft, **D.C.**: Methodology, Validation, **G.Z. and B.H.**: Supervision, **T.M.**: Writing - Review & Editing, **R.B.:** Investigation, Resources, **M.X and X.P.**: Methodology, Investigation, **M.Z. and Y.S.M**: Writing - Review & Editing, Project administration, Funding acquisition, Supervision,

**Declaration of Competing Interest**
The authors declare that they have no known competing financial interests or personal relationships that could have appeared to influence the work reported in this paper.

**Acknowledgements**
The authors acknowledge funding support from the US Department of Energy, Office of Basic Energy Sciences, under award number DE-SC0002357. FIB/SEM was performed at the San Diego Nanotechnology Infrastructure (SDNI), a member of the National Nanotechnology Coordinated Infrastructure, which is supported by the National Science Foundation (Grant ECCS-2025752). TEM was performed at the UC Irvine Materials Research Institute (IMRI). This work used Stampede 2 at Texas Advanced Computing Center (TACC) through allocation DMR110008 from the Extreme Science and Engineering Discovery Environment (XSEDE), which was supported by National Science Foundation grant number #1548562.